\PassOptionsToPackage{square,comma,numbers}{natbib}
\documentclass[12pt]{article}
\usepackage[preprint]{neurips_2023}
\usepackage{amsmath,amsfonts}
\usepackage{float}
\usepackage[pdftex]{graphicx}
\usepackage[T1]{fontenc}
\usepackage[utf8]{inputenc}
\usepackage{siunitx}
\usepackage{textcomp}
\usepackage{xr-hyper} 
\usepackage{hyperref}
\usepackage{amssymb}
\usepackage{amsmath}
\usepackage{tabularray}
\usepackage{booktabs}
\hypersetup{
    colorlinks=true,
    linkcolor=black,
    filecolor=magenta,      
    urlcolor=blue,
    citecolor=black,
}
\makeatletter

\begin{document}

\title{Chaining thoughts and LLMs to learn DNA structural biophysics}

\author{%
  Tyler D.~Ross\\
  Department of Physics\\
  MIT\\
  \texttt{tyler.qed.ross@gmail.com} \\
  \And
  Ashwin Gopinath\\
  Department of Mechanical Engineering\\
  MIT\\
  \texttt{agopi@mit.edu}\\
}

\maketitle


\begin{abstract}
    \noindent The future development of an AI scientist, a tool that is capable of integrating a variety of experimental data and generating testable hypotheses, holds immense potential. So far, bespoke machine learning models have been created to specialize in singular scientific tasks, but otherwise lack the flexibility of a general purpose model. Here, we show that a general purpose large language model, chatGPT 3.5-turbo, can be fine-tuned to learn the structural biophysics of DNA. We find that both fine-tuning models to return chain-of-thought responses and chaining together models fine-tuned for subtasks have an enhanced ability to analyze and design DNA sequences and their structures.

\end{abstract}

\section{Introduction}
With the ever growing accumulation of experimental data across vast fields of science, it has become clear that a machine learning model that is capable of integrating all this data and generating new hypotheses would revolutionize the advancement of scientific knowledge. Specialized models have been created to search for new physics \cite{Karagiorgi2022MLPhysics} or predict the structures of proteins \cite{Jumper2021AlphaFold}. However, the recent success of general purpose large language models (LLMs) have shown that a single model is capable of performing a wide range of tasks, such as programming, translation, and summarization. The textual representation that is typically used to represent amino and nucleic acid polymers appears well-suited for LLM frameworks. In fact, recent studies have shown that the transformer architecture of LLMs can also be utilized in constructing specialized models to predict the structures of biomolecules \cite{LinProteinStructureWithALanguageModel2023, Chowdhury_ProteinLanguageModel2022, li2021bioseqblm}. However, as a small step toward the goal of an interdisciplinary AI scientist, we are interested in determining if a general purpose LLM can be fine-tuned to model a physical phenomenon. Specifically, we aim to test if chatGPT 3.5-turbo can learn the biophysics of DNA structure formation.


The structural biophysics of nucleic acids forms the basis of DNA nanotechnology, which harnesses the predictable nature of DNA base pairing and molecular self-assembly to create complex nanostructures as well as perform computations \cite{Seeman1982NucleicAcidJunctions, Rothemund2006Origami,Ke2012DNABricks,Seelig2006DNALogic}. A key component to DNA nanotechnology is the formation of secondary structures, which are the specific shapes or arrangements that DNA molecules can assume beyond the familiar double helix of the primary structure. Secondary structures are largely governed by two phenomena, the first being Watson-Crick base pairing between DNA's four nucleotides: adenine (A), thymine (T), cytosine (C), and guanine (G). Adenine pairs with thymine (A-T) and cytosine pairs with guanine (C-G) through highly specific hydrogen bonding. Notably, C-G bonds are more energetically favorable than A-T bonds. Base pairing is dependent on the directionality of the DNA strand, denoted as 5' (five-prime) to 3' (three-prime) ends. Reverse complements are sequences that would bind to each other when aligned in opposite directions. For instance, the sequence 5'-AGTC-3' would have a reverse complement of 5'-GACT-3'. Beyond base pairing, there is also an attraction between paired bases in the form of stacking bonds. As with base pairing, the strength of stacking bonds varies as a function of the base pairs involved. The likelihood of a given structure forming at thermodynamic equilibrium is determined by its free energy. The free energy of a DNA structure is typically approximated by a nearest neighbor model that uses the energies of pairwise stacking bonds, which have been determined empirically, plus the energy from base pairing. 

These energetic approximations are the basis of the NUPACK software suite \cite{Zadeh2011NUPACK,fornace_2023_nupack}, which is the standard in the DNA nanotech community for the analysis and design of DNA and RNA structures. NUPACK predicts the most stable structure that will form for a given set of DNA strands by finding the structure that has the minimum free energy (MFE) of all possible structures (up to a user-designated complex size limit). The DNA secondary structures are represented using parens-dot-plus notation, which uses parentheses to indicate base paired regions, dots for unpaired nucleotides, and pluses to separate multiple strands. For example,  ``((((..+....))))'' represents two strands where the first four bases of strand 1 are bound to the last four bases of strand 2. For this work, we will be using NUPACK to provide the data for training and validation. It is worth emphasizing, however, that NUPACK is not a complete biophysical model. For example, psudoknoted structures as well as Hoogsteen base pairing, which allows for the formation of triplexs and quradraplexs, are both ignored. In this work, we will be using a simplified data set, where DNA pairs are of equal length and are not self-complementary (i.e. the strands do not bind to themselves). This allows us to separate out the capability of the LLMs to learn the fundamentals of DNA structural biophysics from the task of searching across the space of all possible configurations.

\section{Methods}
\begin{figure}[hbt!]
    \centering
    \includegraphics[width=\textwidth]{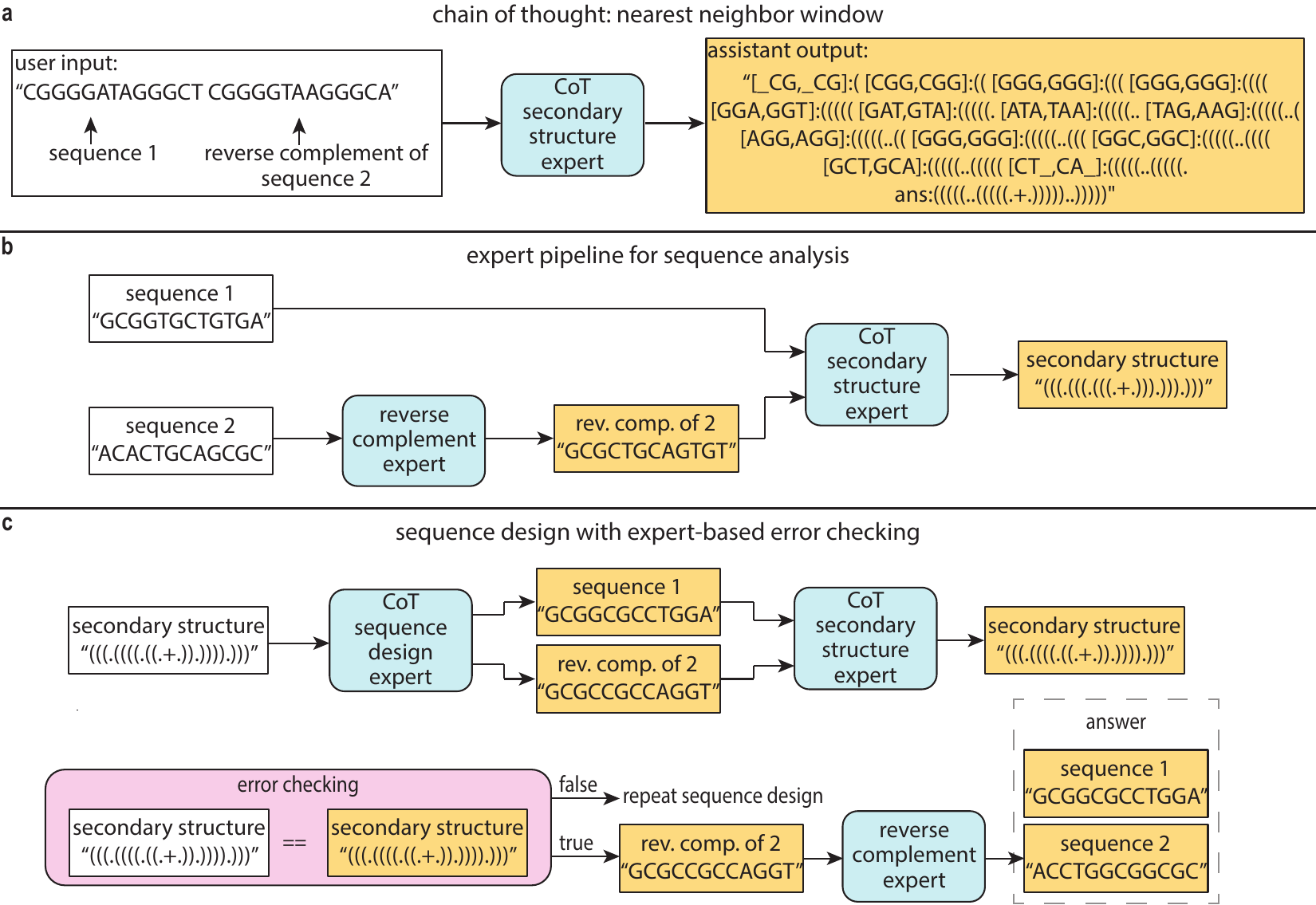}
    \caption{Schemes for chain of thought (CoT) and pipelines of models used to perform sequence analysis and design. \textbf{a}, Chain of thought fine-tuning where the model prints out each base and their neighbors before determining if a stable base pair is formed. \textbf{b}, A sequence analysis pipeline that uses a model that is tuned to provide the reverse complement of a sequence, which is then fed into a model that is tuned to determine the secondary structure. Boxes in white indicate values provided by the user, teal boxes are fine-tuned models, and orange boxes represent final answers from model outputs. \textbf{c}, Expert-based error checking scheme where the sequences designed by one model are analyzed by another to verify that the desired secondary structure is formed.}
    \label{fig:schemes}
\end{figure}

Our work is built on two primary concepts: fine-tuning models to follow a train of thought approach before arriving at an answer, and using a pipeline of models (experts) to solve different aspects of the problem and provide feedback.

For simplicity in implementation, each expert is a fine-tuning of gpt-3.5-turbo-1106 using OpenAI's API. A basic error checking step is applied to the output of each expert, where the model is queried with the same input (up to 20 times) if the model's final answer is not the expected length or contains invalid characters. Models fine-tuned to follow a chain-of-thought (CoT) approach will step-by-step print out the nearest neighbor window of the input while assembling the structure/sequence pair before giving their final answers (Fig.~\ref{fig:schemes}a). In addition to breaking down the problem into simpler steps as shown in \cite{wei2023chainofthought}, our fine-tuned CoT approach leverages a feature of our problem (i.e. that binding stability is based on nearest neighbor interactions) to give the model a hint as to how to process the input. When the nearest neighbor window is applied, the input string(s) are padded with the character ``\_'' to indicate the ends of the sequences, as bases at the edges are less likely to form stable base pairs due to having only one potential stacking bond. For the CoT approach, we only evaluate a model's performance based on its final answer, which comes after ``ans:''.

In all of our experiments we explore the utilization of an expert pipeline where models that have been fine-tuned to perform subtasks feed into each other. Here we are motivated by the idea that a collection of models that are specialized on subtasks can perform better than a single model \cite{jacobs1991MixturesofExperts}. In the simplest case, we have the output of one model provide part of the input to another model (Fig.~\ref{fig:schemes}b). For the more complex problem of sequence design, we introduce an error checking scheme where the sequences generated by one model are analyzed to see if they form the desired secondary structure (Fig.~\ref{fig:schemes}c). If the structures do match then the sequences are accepted.

The ground truth data used to train and validate the models is created using NUPACK. The sequences are designed by NUPACK to ensure that the MFE structure does not have self-complementarity or require alignment. For sequence design, we use NUPACK to determine the ground truth MFE structure formed by the LLM designed sequences. We set NUPACK's model conditions to 20 $^{\circ}$C, 1 M sodium, and with ensemble stacking. Sequences are of lengths between 10-25 bases with sequence pairs being of matching length. Paired strands have a minimum of 1 mismatch up to 30\% of bases mismatched. The training set size is 10,000 and validated against a set of 1,000 sequences. Learning curves are generated over training sets containing 200, 500, 1,500, 3,700, and 10,000 examples. When evaluating the learning curves for the pipelines, all experts are fine-tuned on data sets of the same size. A separate training and validation set is used for the sequence design model, where each entry in the set is a unique structure.

\section{Experiments and results}

\subsection{Predicting secondary structure}\label{Section:structure_analysis}

We begin exploring the capability of a general purpose LLM to learn the secondary structure of DNA by testing four different approaches: naive, chain-of-thought (CoT), reverse complement followed by CoT, and a pipeline of a reverse complement expert feeding into a secondary structure expert. The naive approach is to provide two DNA sequences and have the model return the corresponding secondary structure in dot-parens-plus notation. For example, Input: ``ACCGCGCCCT TGGGCGGGGA'' Output: ``.((.(((((.+.))))).)).''. For CoT, the input remains the same but the output applies a nearest neighbor window as it determines the structure, Input: ``CCCGGCGCTG CTGCGGCGGG'' Output: ``[\_CC,\_GG]:( [CCC,GGG]:(( [CCG,GGC]:((( [CGG,GCG]:(((( [GGC,CGG]:((((. [GCG,GGC]:((((.( [CGC,GCG]:((((.(( [GCT,CGT]:((((.((( [CTG,GTC]:((((.(((. [TG\_,TC\_]:((((.(((.. ans:((((.(((..+..))).))))''. We next vary the CoT approach by breaking the problem down further. The models so far had to recognize that the relationship between the strands is from left to right on the first sequence and from right to left on the second sequence and that further it must learn the pairing rule of G-C and T-A all within one step. We therefore introduce a third condition where the model first produces the reverse complement of the second sequence and then performs a CoT base comparison between the first sequence and the reverse complement of the second sequence. For example, Input: ``GCAGGCCCGAA AACGGGGCTGC'' Output: ``GCAGCCCCGTT [\_GC,\_GC]:( [GCA,GCA]:(( [CAG,CAG]:((( [AGG,AGC]:(((( [GGC,GCC]:((((. [GCC,CCC]:((((.( [CCC,CCC]:((((.(( [CCG,CCG]:((((.((( [CGA,CGT]:((((.(((( [GAA,GTT]:((((.((((. [AA\_,TT\_]:((((.((((.. ans:((((.((((..+..)))).))))''. Finally, we break the problem down even further by creating a pipeline where a separate model is responsible for determining the reverse complement and sends it to another model that determines the secondary structure. For example, first we provide the reverse complement expert with sequence 2: ``GGCCGCCGAC'' and receive its reverse complement ``GTCGGCGGCC'', we then give the secondary structure expert the Input: ``GTCGGCGGGC GTCGGCGGCC'' Output: ``[\_GT,\_GT]:( [GTC,GTC]:(( [TCG,TCG]:((( [CGG,CGG]:(((( [GGC,GGC]:((((( [GCG,GCG]:(((((( [CGG,CGG]:((((((( [GGG,GGC]:(((((((( [GGC,GCC]:((((((((. [GC\_,CC\_]:((((((((.( ans:((((((((.(+).))))))))''. For comparison, we also run this pipeline where a reverse complement algorithm provides the ground truth reverse complement to the secondary structure expert.

\begin{table}[htb]
\centering
\begin{tabular}{@{}cccccc@{}}
\toprule
            & naive & CoT  & \begin{tabular}[c]{@{}c@{}}rev. comp.\\ \& CoT\end{tabular} & \begin{tabular}[c]{@{}c@{}}pipeline\\ rev. comp. expert\\ $\rightarrow$ CoT\end{tabular} & \begin{tabular}[c]{@{}c@{}}pipeline\\ ground truth\\ rev. comp.\\ $\rightarrow$ CoT\end{tabular} \\ \midrule
accuracy \% & 7.4   & 77.4 & 88.9  & 92.8 & 95.4  \\ \bottomrule
\end{tabular}
\vspace{10pt}\caption{Secondary structure prediction accuracy.}
\label{table:secondary_structure}
\end{table}

 \begin{figure}[hbt!]
    \centering
    \includegraphics[width=0.75\textwidth]{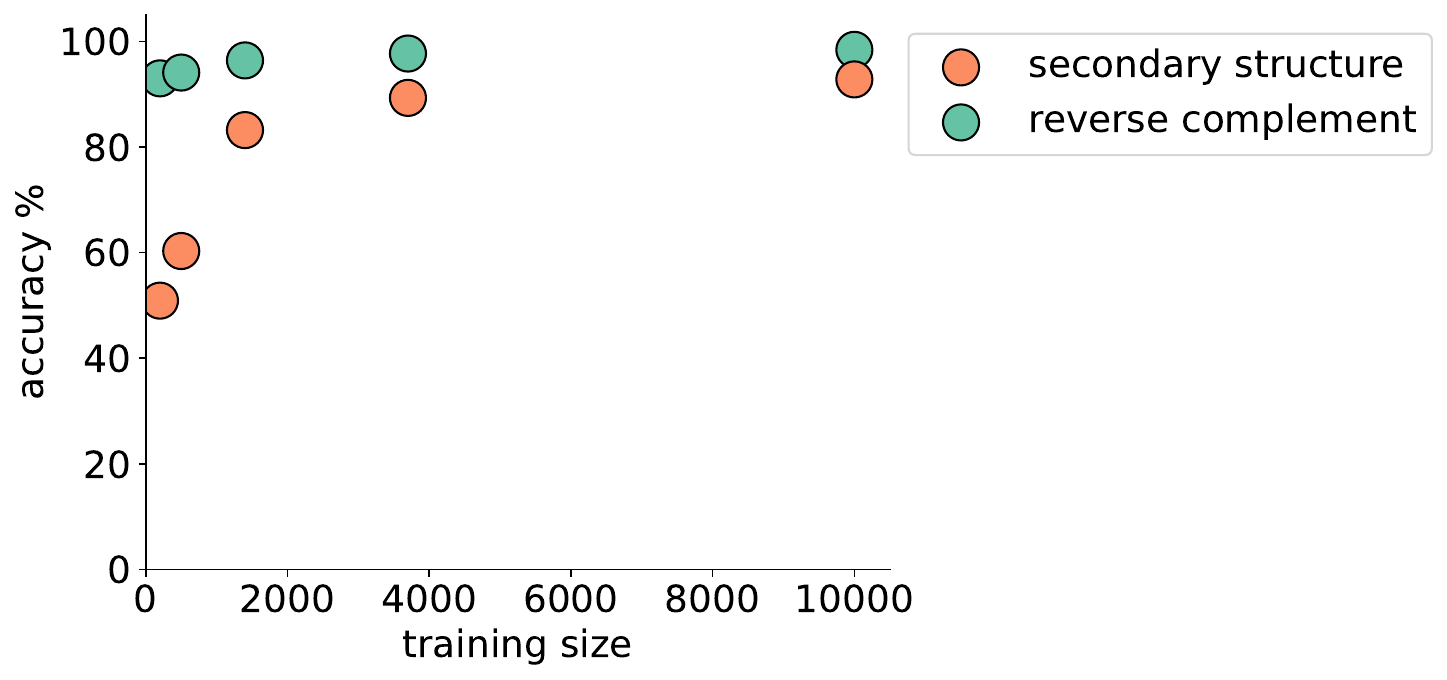}
    \caption{Learning curves for secondary structure prediction and the reverse complement expert. For the secondary structure, we are using the expert pipeline approach.}
    \label{fig:structure_learning}
\end{figure}

\textbf{Results}~~~The general trend we find is that the more we break down the problem the better the performance (Table~\ref{table:secondary_structure}). In all cases the majority of errors are due to incorrect base pairing rather than issues with formatting or incorrect structure lengths. Having the model explicitly write out the nearest neighbor window results in a substantial improvement over the naive approach. The benefit of CoT here presumably comes from both expanding the context before the answer is given and, more specifically, by breaking down the relevant character relationships. It is perhaps surprising how large an improvement we see when the model produces the reverse complement of the second sequence before performing CoT. The comparison of one sequence from left-to-right with another sequence from right-to-left may be inherently difficult for the model to learn given the unidirectional nature of the transformer decoder architecture. Producing the reverse complement first instead allows the model to compare both strings from left to right. For the pipeline approach, we additionally fine-tune a model to return the reverse complement of a sequence, which has an accuracy of 98.4$\%$. We note that, unlike the determination of the secondary structure, finding the reverse complement of a sequence follows a simple closed-form function. Perhaps this is why the reverse complement expert saturates at small training sizes in comparison to the secondary structure model (Fig.~\ref{fig:structure_learning}). The difference in the performance between the expert pipeline and the ground truth pipeline suggests that there is largely no overlap in where errors occur for the structure and reverse complement experts. Therefore, there may be a diminishing return in terms of the number of experts a pipeline uses. Ultimately, offloading the reverse complement operation to another model enhances the accuracy of the secondary structure prediction.

\subsection{Calculating minimum free energy}

 To further test the model's ability to learn the structural physics of DNA, we ask it to calculate the minimum free energy (MFE) for a pair of sequences. As mentioned in the Introduction, the free energy is approximated as a function of both the base pairing and the nearest-neighbor stacking bond energies, which are empirically determined. Here we challenge the model to find the MFE without explicitly writing out any mathematical operations. We explore MFE calculation across four approaches: naive, reverse complement followed by CoT, a pipeline where a model provides the reverse complement to a CoT MFE model, and a variation on the naive approach where the model is provided with both the ground truth reverse complement and ground truth secondary structure. The first three cases are nearly identical to what is described in the previous section, but on the last step the model returns the MFE instead of the structure. Also as before, we test the pipeline method where the reverse complement expert is substituted for a ground truth algorithm. An example of the ground truth reverse complement and structure approach is as follows, Input: ``CGTTTTTCCGACTTGCGCCG CGAATTTCCGACGTGCGCCG ((..((((((((.(((((((+))))))).))))))))..))'' Output: ``-28.3''. The idea here is to test if providing the model with the secondary structure as part of the input is sufficient to match the model's performance over the CoT approach.
 
\begin{table}[htb]
\centering
\begin{tabular}{@{}cccccc@{}}
\toprule
            & naive & \begin{tabular}[c]{@{}c@{}}rev. comp. \\ \& CoT\end{tabular} & \begin{tabular}[c]{@{}c@{}}pipeline\\ rev. comp.  expert\\ $\rightarrow$ CoT\end{tabular} & \begin{tabular}[c]{@{}c@{}}pipeline\\ ground truth\\ rev. comp.\\ $\rightarrow$ CoT\end{tabular} & \begin{tabular}[c]{@{}c@{}}ground truth\\ rev. comp.\\ \& ground truth\\  structure\end{tabular} \\ \midrule
error $\left( \frac{\text{kcal}}{\text{mol}} \right)$ & 1.67 $\pm$ 1.43 & 1.55 $\pm$ 1.83 & 1.23 $\pm$ 1.63 & 1.15 $\pm$ 1.26& 1.43 $\pm$ 1.19 \\ \bottomrule
\end{tabular}
\vspace{10pt}\caption{Minimum free energy error from sequence analysis. Error is mean absolute error and the $\pm$ term represents standard deviation of the absolute error.}
\label{table:MFE}
\end{table}

 \begin{figure}[hbt!]
    \centering
    \includegraphics[width=0.45\textwidth]{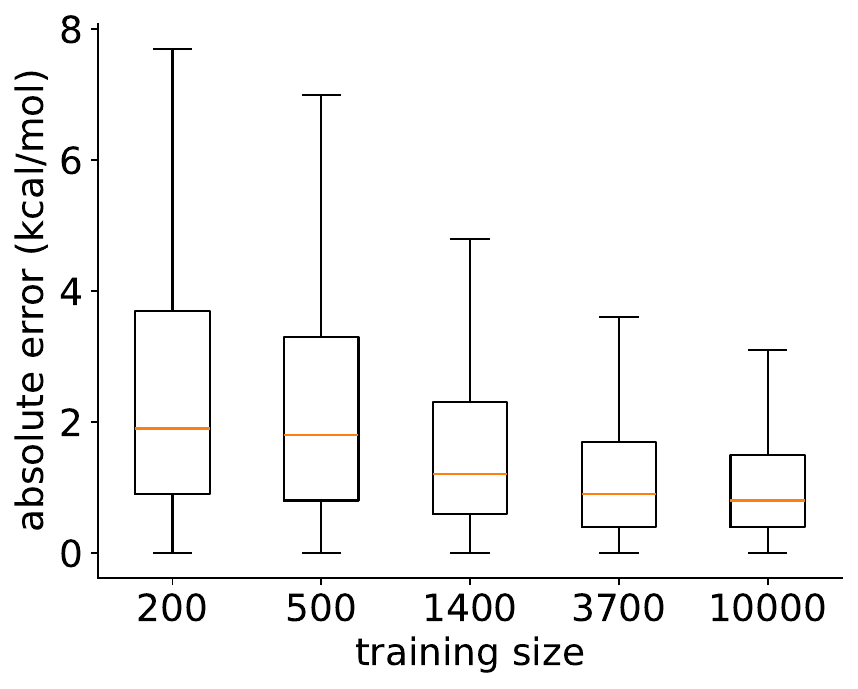}
    \caption{Impact of training set size on MFE predictions for the reverse complement expert pipeline approach.}
    \label{fig:MFE_learning}
\end{figure}
 
 \textbf{Results}~~~The model appears to have learned some aspects of determining the MFE (Table~\ref{table:MFE}). As a point of reference, the standard deviation of the validation set is 6.04 kcal/mol. LLMs generally struggle with mathematical tasks \cite{frieder2024mathematical}, so it is not unexpected that we do not see as large an improvement between the various approaches as we did in the previous section. With that said, increasing the amount of training data showed a significant improvement in accuracy (Fig.~\ref{fig:MFE_learning}). Like the previous section, it appears that the combination of the CoT and model pipeline provides the best result. It is also worth highlighting that the CoT process results in a lower error than providing the model with the ground truth secondary structure. This suggests that the context provided by the nearest neighbor window is not just a means to finding the secondary structure but is, to some extent, guiding the model to learn the base pairing/stacking energies.

\subsection{Sequence design with expert error checking}
The design problem, creating DNA sequences that will form a desired secondary structure, is considerably more difficult than the analysis problems we have attempted so far. Slight errors in the sequence design can result in self-complementarity or alignment issues. Once again, we apply a naive, CoT, and pipeline approach to solving the problem. For the naive approach, we provide the secondary structure and have the model return a pair of sequences. For example, Input: ``(((((..((((((((((.(+).))))))))))..)))))'' Output: ``GCGTTCATGTCTTGCGGGC GGCCGCAAGACACAAACGC''. When applying the CoT approach, we have the model apply a nearest neighbor window to the secondary structure as it builds out the sequences. Based on the result from Section~\ref{Section:structure_analysis}, have the CoT build out the second sequence as its reverse complement and then on the final step transform it into its forward sequence. For example, Input: ``..((.(((.(((+))).))).))..'' Output: ``[\_..]:[A,T] [..(]:[AA,TT] [.((]:[AAC,TTC] [((.]:[AACA,TTCA] [(.(]:[AACAG,TTCAC] [.((]:[AACAGC,TTCACC] [(((]:[AACAGCC,TTCACCC] [((.]:[AACAGCCG,TTCACCCG] [(.(]:[AACAGCCGG,TTCACCCGC] [.((]:[AACAGCCGGC,TTCACCCGCC] [(((]:[AACAGCCGGCC,TTCACCCGCCC] [((\_]:[AACAGCCGGCCG,TTCACCCGCCCG] ans:AACAGCCGGCCG CGGGCGGGTGAA''. In the third condition we have a pipeline where the reverse complement transformation on the final step is offloaded to a second model or ground truth algorithm.

We now add in an error checking layer into our pipeline of experts (Fig.~\ref{fig:schemes}c). Once our design model produces a sequence pair, we have the expert that we tuned to predict secondary structures (Section~\ref{Section:structure_analysis}) to verify that sequences form the desired structure. If the structures are found to be the same, then the second strand is transformed into its forward sequence and the final sequence pair is returned. If the structures are found to be different, then the design process is repeated again for the same input structure. For comparison, we also test this approach by having a ground truth error check where NUPACK performs the sequence analysis in the error checking step.

\begin{table}[htb]
\centering
\begin{tabular}{@{}ccccc@{}}
\toprule
                     & naive & \begin{tabular}[c]{@{}c@{}}CoT\\ \& rev. comp.\end{tabular} & \begin{tabular}[c]{@{}c@{}}pipeline\\ CoT\\ $\rightarrow$ rev. comp. expert\end{tabular} & \begin{tabular}[c]{@{}c@{}}pipeline\\CoT\\ $\rightarrow$ ground truth rev. comp.\end{tabular} \\ \midrule
accuracy \%          & 3.9                                     & 77.6                                                                                          & 85.9                                                                          & 88.9                                                                      \\ \midrule
\multicolumn{1}{l}{} & \multicolumn{2}{c}{\begin{tabular}[c]{@{}c@{}}pipeline\\CoT\\ $\rightarrow$ expert error checking \\ $\rightarrow$ rev. comp. expert\end{tabular}} & \multicolumn{2}{c}{\begin{tabular}[c]{@{}c@{}}pipeline\\CoT\\ $\rightarrow$ ground truth error checking\\ $\rightarrow$ rev. comp. expert\end{tabular}}                       \\ \midrule
accuracy \%          & \multicolumn{2}{c}{93.1}                                                                                                                & \multicolumn{2}{c}{99.8}                                                                                                                                  \\ \bottomrule
\end{tabular}
\vspace{10pt}\caption{Sequence design accuracy.}
\label{table:design}
\end{table}

 \begin{figure}[hbt!]
    \centering
    \includegraphics[width=0.5\textwidth]{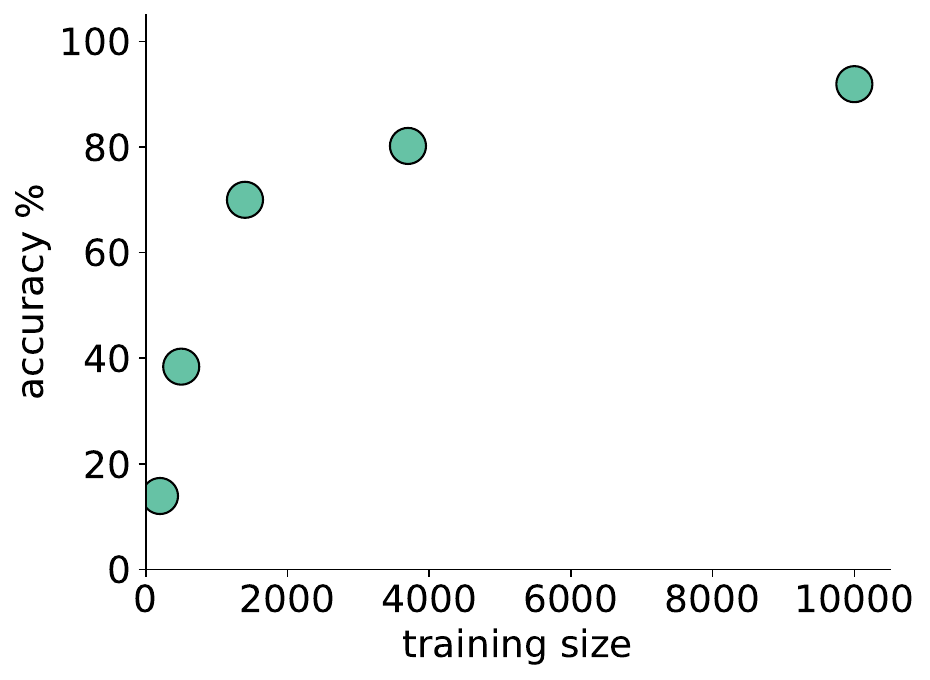}
    \caption{Sequence design learning curve for the pipeline with expert error checking approach.}
    \label{fig:design_learning}
\end{figure}

\textbf{Results}~~~In spite of the analysis and design tasks being quite different from each other, we see similar benefits from the same approaches. With that said, comparing the results of the first row of approaches in Table~\ref{table:design} to their analogs in Table~\ref{table:secondary_structure} confirms that the model has greater difficulty in performing sequence design. The addition of the error checking layer, however, appears to compensate for the increase in difficulty. In fact, we see that with ground truth error checking, the designer can perform almost perfectly. Based on the steepness of the learning curve, design has more benefit from larger training sets than analysis does (Fig.~\ref{fig:design_learning}). For generating this particular learning curve, we reduced the number of retries per error to 3 as it would otherwise run for several days for the validation of the low training sizes to run. The accuracy at the training size of 10,000, is similar between 3 retires and 20 retries. Provided how successful our simple error handling scheme is, it is also worth considering if error correction may be implemented. For example, a model that is fine-tuned to reflect on errors and produce corrections may have even better performance.

\section{Conclusion}
We have shown that a general purpose LLM can be fine-tuned to learn the structural biophysics of DNA. This is made possible through the combination of chain-of-thought responses and the chaining of experts. In particular, breaking down a complicated task into a series of subtasks that models can master and chaining those models together may be a useful concept for constructing more powerful models that are capable of working on scientific problems.

One direction in which this work can be further taken is to test if smaller models can be chained together for a similar performance improvement. The implication being that models requiring lower compute resources could collectively perform more complex tasks than a singular large model. For this particular type of problem, an LLM architecture involving both an encoder and decoder may perform better at direct sequence comparison than a decoder-only architecture.

As noted in the Introduction, our datasets are within a simpler subspace of the complete range of possible DNA interactions. One may be able to make a more generalized model by using CoT to search through configurations to deal with self-complementarity and sequence alignment. We also highlight that the concepts from this work will also apply to RNA structures. From the perspective of DNA nanotech, there is a potential opportunity for LLMs to eventually be useful where NUPACK fails. For example, an LLM may be capable of learning to predict pseudoknots, Hoogsteen base pairing, and many-strand interactions (which has utility for designing optimal staple strands for DNA origami). Further, an LLM trained on experimental data may produce more accurate structural predictions than the approximations used by current models.

\section{Code availability}
All code and data used is available at the following link: \href{https://github.com/TDRoss/DNA-LLM}{https://github.com/TDRoss/DNA-LLM}.

\bibliography{zotero.bib}

\end{document}